\documentclass[aps,prl,preprint,showpacs]{revtex4}

\usepackage{graphicx}

\begin{document}

\title{Experimental evidence for a collective insulating state in two-dimensional superconductors} 

\author{G. Sambandamurthy$^1$} 
\author{L. W. Engel$^2$} 
\author{A. Johansson$^1$}
\author{ E. Peled$^1$}
\author{D. Shahar$^1$}
\affiliation{$^1$Department of Condensed Matter Physics, Weizmann Institute of Science, Rehovot 76100, Israel, \\ $^2$National High Magnetic Field Laboratory, Florida State University, Tallahassee, Florida 32306, USA}

\date{\today}

\begin{abstract}
We present the results of an experimental study of the current-voltage characteristics in strong magnetic field ($B$) of disordered, superconducting, thin-films of amorphous Indium-Oxide. As the $B$ strength is increased superconductivity degrades, until a critical field ($B_c$) where the system is forced into an insulating state. We show that the differential conductance measured in the insulating phase vanishes abruptly below a well-defined temperature, resulting in a clear threshold for conduction. Our results indicate that a new collective state emerges in two-dimensional superconductors at high $B$. 
\end{abstract}

\pacs{74.25.Fy, 74.78.-w, 74.25.Dw, 73.61.Ng}

\maketitle

When a superconductor is cooled below a characteristic temperature, $T_c$, its electrical resistivity abruptly vanishes. This phenomenon is a manifestation of the formation of a quantum-mechanical, collective, many-body state of electron pairs \cite{Tinkham96}. Application of a magnetic field ($B$) causes a reduction of $T_c$ until a critical field ($B_c$) where the zero-resistivity state is completely destroyed. In the two-dimensional (2D) limit, evidence indicates that the system is then forced into an insulating state. In this case, $B_c$ is identified with the critical field of the superconductor-insulator transition \cite{Fisher90,Fisher91,Hebard90,Zant92,Yazdani95,Markovic98_1,Gantmakher00_2,Murthy04}. The superconductor-insulator transition (SIT) in 2D systems is theoretically considered within the framework of continuous quantum phase transitions \cite{Sondhi97}. In this framework, the $B$-driven transition reflects a change in the macroscopic state, rather than the demise, of the Bose-like Cooper pairs that underlie the superconducting phase. Fisher and co-workers \cite{Fisher90,Fisher91} postulated a dual description of the SIT in which the superconducting and the insulating phases are caused by condensations of Cooper-pairs and vortices, respectively, into a superfluid state. While several experimental \cite{Hebard90,Zant92,Yazdani95,Markovic98_1,Gantmakher00_2,Murthy04,Christiansen02} and theoretical \cite{Fisher90,Fisher91} studies are supportive of this conjecture, others \cite{Valles92} correlate the SIT with the vanishing of the superconducting gap, indicating that Cooper-pairs do not survive the transition to the insulator.  In this Letter, we present results that indicate that a new collective state emerges in the $B$-induced insulating phase. 

Our data were obtained from studies of disordered thin films of amorphous Indium-Oxide (a:InO). The films were prepared by e-gun evaporating high purity (99.999 \%) In$_2$O$_3$ on clean glass substrates in a high vacuum system. The thickness was measured in situ by a Quartz crystal thickness monitor. The thickness of all the films used in the present study was 300 $\rm \AA$. The films were examined by Atomic Force Microscopy and the surface images show that the films were continuous without any voids. In order to learn more about the microstructure of the films, we performed transmission electron microscopy studies on films that were prepared under the same conditions as the samples used for transport measurements. Diffraction patterns and micrographs show that the films were amorphous and crystalline inclusions were never observed. For transport measurements, the films were lithographically defined to Hall-bar patterns, with voltage probe separation twice the width of the Hall-bar. The widths of the samples used in this study are 10-100 $\mu$m. 

Four-terminal resistance measurements were carried out by low frequency (3-13 Hz) AC lock-in techniques, with excitation currents of 1-10 nA. When resistance values were high and conventional four-probe AC lock-in measurements were not possible, we have performed two-probe resistance measurements with an excitation voltage of 10 $\mu$V. Differential resistance (four-probe) and conductance (two-probe) measurements were performed using low frequency AC lock-in techniques combined with DC current and voltage excitations, respectively. The samples were cooled in a dilution refrigerator with a base temperature of 0.01 K. Magnetic fields up to 12 T were applied perpendicular to the surface of the film.

Recent experiments provided evidence that at least some of the superconducting correlations remain in the insulating state of amorphous Indium-Oxide films \cite{Gantmakher00_2,Murthy04,Paalanen92,Gantmakher95}. The insulating behaviour showed a non-monotonic dependence on applied $B$, resulting in a resistance peak with a typical activation energy \cite{Murthy04} that is close to the superconducting transition temperature of the film at $B$ = 0. Such behaviour was argued to result from the presence of localized pockets of superconductivity that still participate in the transport process in the insulating phase.

The $B$-induced insulating peak can be seen in Fig. 1(a), where we plot the sheet resistance ($\rho$) of one of our superconducting films (ABa1) as a function of $B$ at four different temperatures ($T$) ranging from 0.01-0.4 K. The ordinate is presented in a log-scale to span the large range of $\rho$. We observe the transition from the superconducting to the insulating phase, identified with the crossing of the various $\rho$-isotherms, at $B_c$ = 1.72 T. For $B < B_c$, superconductivity prevails and $\rho$ vanishes as $B$ and $T$ tend to zero. When $B$ is increased beyond $B_c$, insulating behaviour sets-in and $\rho$ increases strongly with decreasing $T$. The insulating phase exhibits a non-monotonic behaviour with the $\rho$ traces going through a maximum centred around 9 T, where the insulator is strongest. 

The main purpose of this Letter is to present an intriguing behaviour we observe in the current-voltage characteristics of the $B$-induced insulating phase discussed above. Before we do that it is instructive to examine the evolution of the non-linear current-voltage characteristics from the superconducting state through the transition and into the insulator. In Fig. 1(b) we plot the four-terminal differential resistance ($dV/dI$), as a function of the DC current ($I_{dc}$), taken from the film of Fig. 1(a), measured at our base $T$ = 0.01 K. We show three representative traces that are measured in the $B$ = 0.5-2 T range.  The bottom trace in Fig. 1(b), taken at $B$ = 0.5 T, is typical of a superconductor: $dV/dI$ is immeasurably low as long as $I_{dc}$ is below a well-defined value, $I_{dc}^c$, which is the critical current of the superconductor at that particular $B$-value. When $I_{dc} > I_{dc}^c$, superconductivity is destroyed and a dissipative state emerges. The situation is different for the middle trace of Fig. 1(b) taken at $B$ = 1.5 T. Even at $I_{dc}$ = 0 a zero-resistance state is not observed down to our lowest $T$. However, the current-voltage characteristic still maintains a superconducting flavor: it is non-Ohmic and $dV/dI$ increases with increasing $I_{dc}$. We therefore consider the $B$ = 1.5 T trace to represent a transitional state in the path the system takes from being a superconductor to an insulator. It is a matter of some debate whether this transitional state will develop into a full superconductor in the $T$ = 0 limit \cite{Yazdani95,Phillips03,Ephron96,Chervenak00}. Experimental realization of this limit is not possible.

While the bottom two traces in Fig. 1(b) exhibit superconducting traits, the top trace taken at $B$ = 2 T clearly does not. Instead, it has the opposite low-$I_{dc}$ dependence indicative of an insulating state: it is again non-Ohmic, but this time an increase in $I_{dc}$ results in a decrease of $dV/dI$. This is consistent with the $\rho$-B data of Fig. 1(a) where the transition to the insulating phase occurs at $B$ = 1.72 T. While the non-linear current-voltage characteristic deep in the superconducting phase is usually attributed to current-induced vortex depinning \cite{Rzchowski90}, the origin of non-linearity in the insulating phase is less clear, and will be discussed in a future publication.

We now turn to the main result of our study: the current-voltage characteristics of the insulator at $B$-values much higher than $B_c$. Since we are now focusing on the insulator, it is natural to consider the differential conductance ($dI/dV$) rather than $dV/dI$. In Fig. 2 we plot two-terminal $dI/dV$ traces of another sample, Ja5, against the applied DC voltage ($V_{dc}$) measured at $B$ = 2 T, well above its $B_c$ (= 0.4 T). In this Figure we contrast two traces that are measured at $T$ = 0.15 K and 0.01 K. The data taken at $T$ = 0.15 K are typical of an insulator, i.e., they are strongly non-Ohmic, having a low, but measurable, value at $V_{dc}$ = 0, which increases smoothly with increasing $V_{dc}$. No clear conduction threshold can be identified at this $T$. The response drastically changes when the film is cooled to 0.01 K: as long as $V_{dc}$ is below a well-defined threshold value ($V_T$), the value of $dI/dV$ is immeasurably low. At that $V_{dc}$ (= 4.65 mV for this sample) $dI/dV$ increases abruptly, by several orders of magnitude, and remains finite for higher values of $V_{dc}$. This clearly shows the emergence, at low $T$, of finite $V_T$ for conduction in the insulating phase.

To complement the comparison between the traces measured at two $T$'s, we measured the $T$-dependence of $dI/dV$ in the insulating phase. In Fig. 3 we plot four $dI/dV$ traces, measured at various $V_{dc}$ values, as functions of $T$. All the traces were measured at $B$ = 2 T, in the insulating phase of sample Ja7, whose $B_c$ = 0.3 T. The bottom trace in Fig. 3 is measured with no $V_{dc}$ applied: as $T$ is reduced from 0.2 K, $dI/dV$ decreases smoothly until $T$ $\sim$ 0.08 K, below which the signal disappears in the noise. 

The $T$-dependence is drastically changed for the two traces in the middle, measured with $V_{dc}$ values close $V_T$ (= 1.02 mV for this sample). In the second trace from the bottom, measured at $V_{dc}$ = 1.02 mV, when $T$ is reduced from 0.2 K, $dI/dV$ decreases smoothly until a well-defined temperature, $T^*$ = 0.06 K, where $dI/dV$ drops sharply to immeasurably low values. The third trace from bottom, measured at $V_{dc}$ = 1.07 mV, shows non-monotonic behaviour in addition to the sharp features in $dI/dV$. In the top trace, measured at $V_{dc}$ = 1.5 mV, the $T$-dependence is weak and no abrupt drop in $dI/dV$ is observed down to the lowest $T$.

We have so far described our observation of the emergence, at a well-defined temperature, of sharp thresholds for conduction on the insulating side of the SIT in our films. Before the implications of this observation are discussed, it is important to construct a map of the $B$-driven insulating phase in the $B$--$V_{dc}$ parameter space available in our experiments. In Fig. 4 the behaviour of our films is summarized in the form of a 2D map of the $dI/dV$ values in the $B$--$V_{dc}$ plane for sample Ja5 of Fig. 2. This map was constructed by measuring $dI/dV$ as a function of $V_{dc}$ at $B$ intervals of 0.2 T at 0.01 K. The colours in the map represent the values of $dI/dV$, changing from $dI/dV$ = 0 (dark blue) to $dI/dV$ = $10^{-5}$ $\Omega^{-1}$ (white). The horizontal dashed line marks $B_c$ (= 0.4 T) of this sample.

Sharp conduction thresholds in the insulating phase can be seen in the map as a sudden change in colour in the insulating regime between $B_c$ = 0.4 T and $\sim$ 5 T. As $B$ is increased the threshold behaviour appears at higher values of $V_{dc}$. This trend continues until $B$ is close to the $\rho$-peak, which is at $B$ = 6 T for this sample. Near the $\rho$-peak and beyond it, the $dI/dV$ traces no longer exhibit the sharp thresholds for conduction and $dI/dV$ increases smoothly with $V_{dc}$. This manifests as a gradual change in colours as $V_{dc}$ is changed for $B \geq$ 6 T. 

We are now ready for the central conjecture of our study: the sharp drop in $dI/dV$ values, at a well-defined $T$, is consistent with the condensation of individual charge carriers into a collective state \cite{Fisher90,Christiansen02}. This brings about interesting analogies with a diverse class of physical systems showing similar threshold for conduction that are considered as signatures of collective phenomena. We recall two such examples here: the first is in one-dimensional organic and inorganic solids where thresholds to conduction have been treated as the signature of the depinning of charge density waves \cite{Gorkov89}. Similarly, in 2D electron systems confined in semiconductor heterostructures, thresholds for conduction have been attributed to the depinning of a magnetically induced electron solid akin to the Wigner crystal \cite{Goldman90,Jiang91,Williams91}. 

While the appearance, in our disordered superconducting samples, of a clear threshold voltage for conduction in the insulating phase can be considered as evidence for vortices condensing into a collective state at a well-defined temperature \cite{Fisher90}, we can not rule out the possibility that another physical mechanism, perhaps involving the electrons spins, play a dominant role in forming the threshold behavior. We are currently conducting angle-dependence studies that will illuminate this important point. Regardless of the mechanism involved, we can use the value of $T^*$ to roughly
estimate the size $L$ of the collective region in our samples through the relation $eEL = k_{B}T^*$, where $e$ is the charge of an electron, $E$ is the electric field applied ($E$ = $V_T/d$ with $d$ the voltage probe separation) and $k_B$ is BoltzmannÕs constant. This estimate results in $L \sim$ 80-700 nm, depending on the sample and $B$ value. Our observation of the existence of conduction thresholds in a limited range of $B$ in the insulating phase implies that the collective state does not survive increasing $B$. 

A natural question arising from our observation is whether this unusual threshold behavior is particular to a:InO or is a more general phenomenon associated with disordered superconductors. While we can not answer this question with certainty, we wish to point out the following features of our study  that distinguishes it from earlier studies on disordered superconductors. First, elemental superconductors tend to form granular or polycrystalline structures whereas a:InO films are known to form uniform, amorphous structures. Second, conventional transport measurements performed with a current of 1 nA or above are already beyond the threshold regime observed in our samples. And third, the observation of this novel threshold behavior requires very low temperatures (below 100 mK). 

The differential conductance measurements presented in this Letter were obtained using two-terminal configuration. This is because the very high impedance of the samples prohibited the use of a current source that is compatible with the low-temperature environment. The sample's impedance is so high that even using two-terminal measurements we are only able to establish its lower-bound, around 100 G$\Omega$, and we believe that the actual impedance is much greater. Commercially available current sources that are compatible with low temperature environment have output impedance that are of order 10$^{12}$ -- 10$^{14}$ $\Omega$. Using four-terminal configuration for the $dI/dV$mesurements resulted in instabilities and hysteresis, whenever the sample's impedance became comparable to the output impedance of the current source, rendering it unable to deliver the current through the sample. When the sample's resistance was low, both two and four-terminal measurements yielded virtually the same results. Our results are reproducible between contact configurations and samples indicating that the contacts play a minor, if any, role in the data.

We thank Z. Ovadyahu and Y. Oreg for discussions. This work is supported by the ISF, the Koshland Fund and the Minerva Foundation.

\begin{figure}
\includegraphics{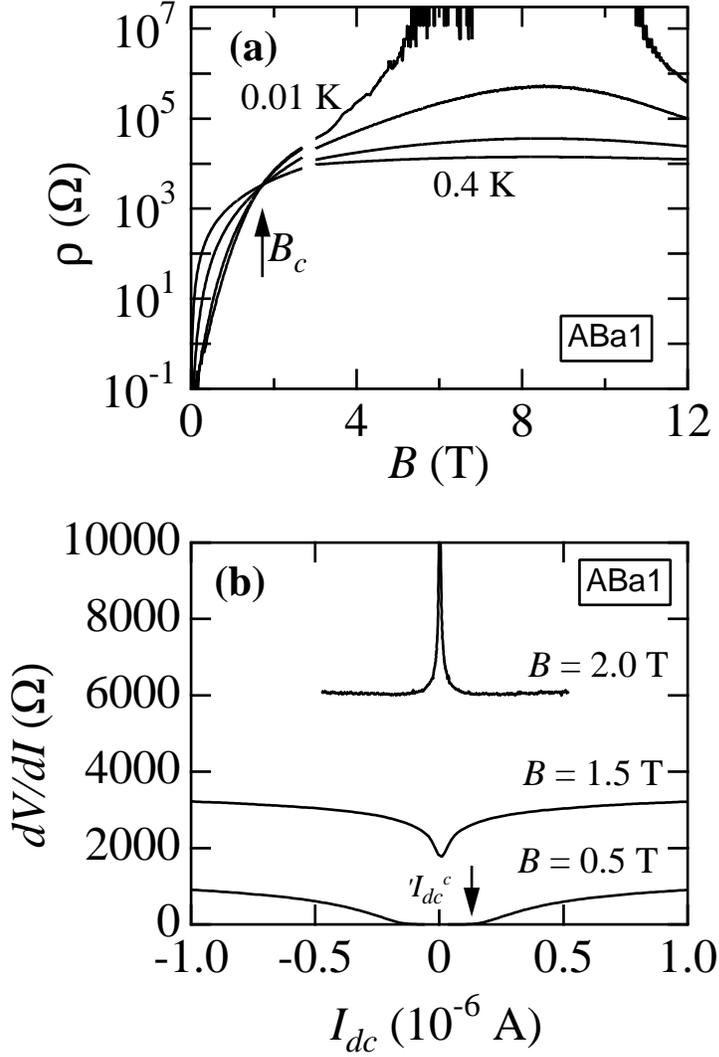}
\caption{(a) Sheet resistance ($\rho$) as a function of $B$-field, of the sample ABa1, measured at four different $T$'s=0.01 K, 0.1 K, 0.2 K and 0.4 K. In the low $B$ range we have plotted four-terminal $\rho$ data measured with AC excitaion currents of 1 nA. In the higher $B$ range (above 3 T) we have plotted two-terminal $\rho$ data measured with AC excitation voltages of 10 $\mu$V. The vertical arrow marks $B_c$ (= 1.72 T), the critical field of the $B$-driven SIT. (b) Evolution of the current-voltage characteristics across the superconductor-insulator transition.  Four-terminal differential resistance, of the same film in (a), is plotted against the DC current for $B$ values of 0.5, 1.5 and 2.0 T. All traces are measured at $T$ = 0.01 K with AC excitation currents of 10 nA.}
\end{figure}

\begin{figure}
\includegraphics{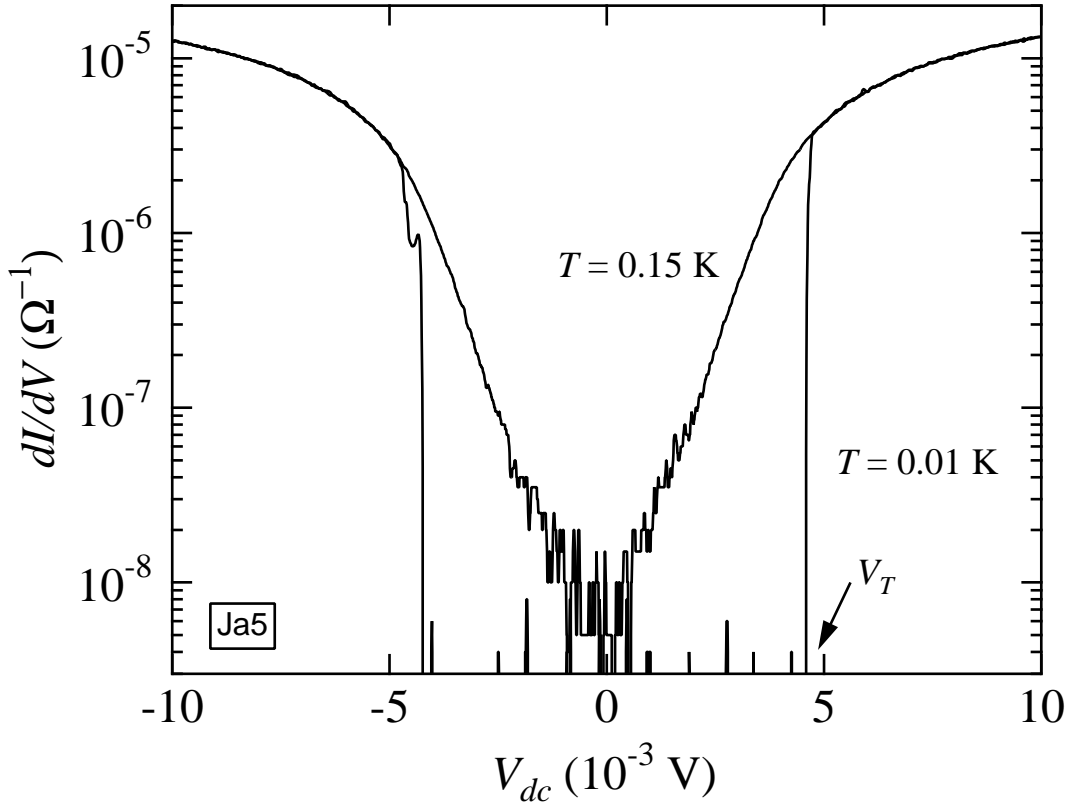}
\caption{Comparison of the current-voltage characteristics of the $B$-driven insulating phase at two $T$Õs (0.15 K and 0.01 K). The traces show the two-terminal differential conductance measured at $B$ = 2 T as a function of DC Voltage. The AC excitation voltage applied is 10 $\mu$V. The sample used is Ja5 with $B_c$ = 0.4 T. $V_T$ marks the threshold voltage for conduction at $T$= 0.01 K.}
\end{figure}

\begin{figure}
\includegraphics{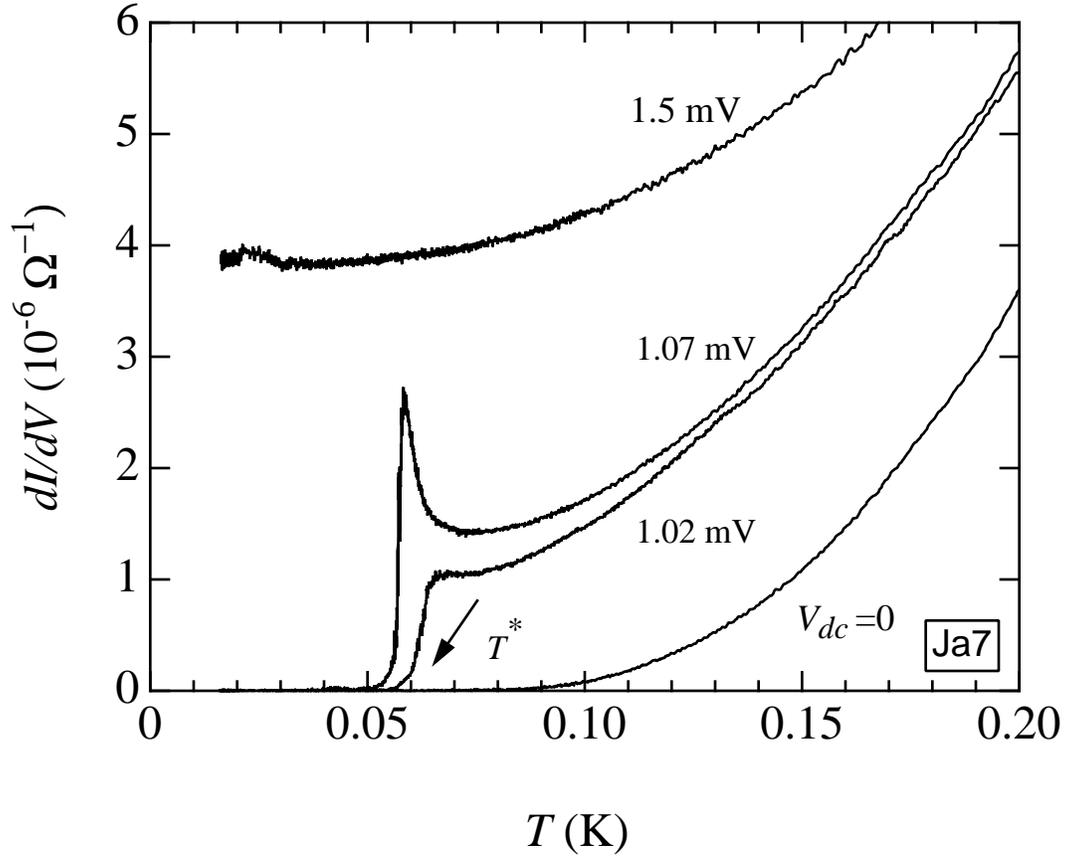}
\caption{Temperature dependence of $dI/dV$ in the $B$-induced insulating phase. $dI/dV$ traces, measured at $B$ = 2 T, of the sample Ja7 with $B_c$ = 0.3 T. The AC excitation voltage used is 10 $\mu$V. The threshold voltage ($V_T$) for this sample $\sim$ 1.02 V. The $V_{dc}$ values used are (from bottom to top): 0, 1.02, 1.07 and 1.5 mV. $T^*$ is temperature at which $dI/dV$ vanishes abruptly.}
\end{figure}

\begin{figure}
\includegraphics{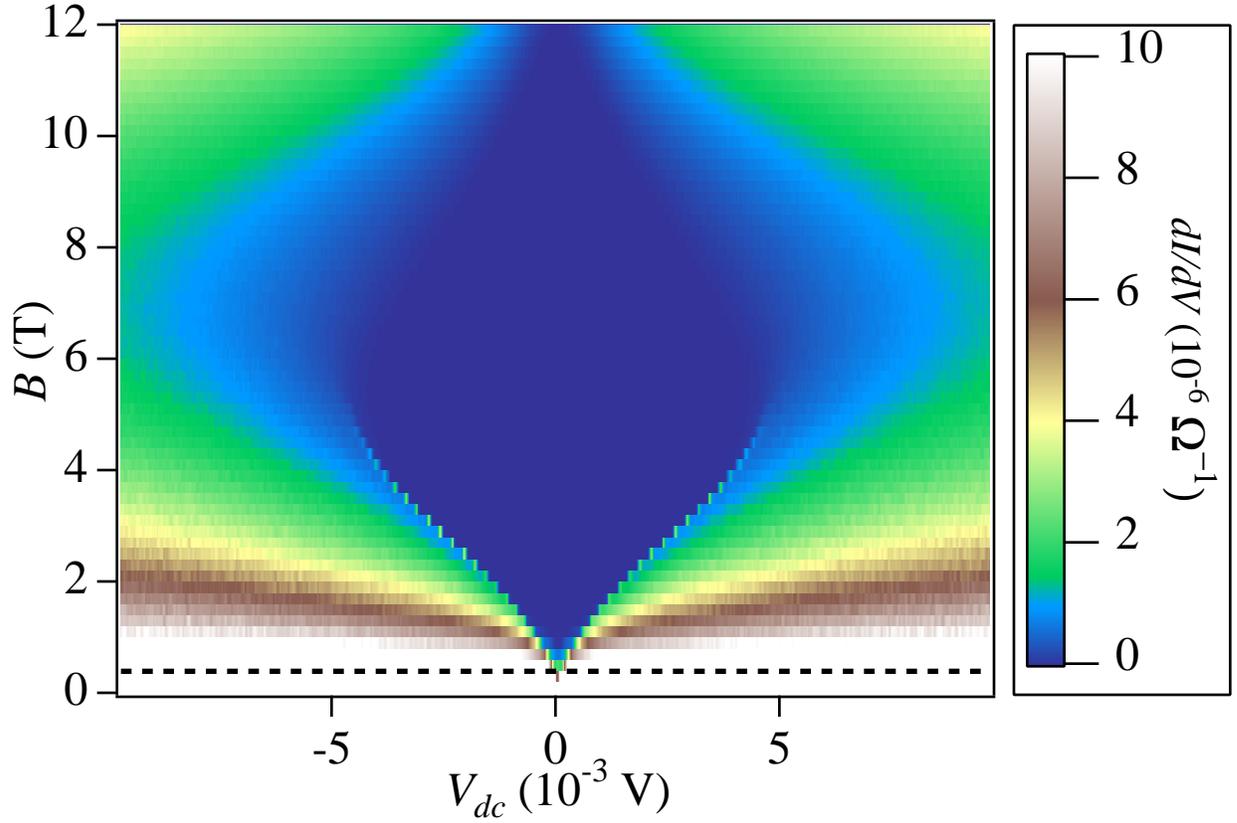}
\caption{Two-dimensional map of the $dI/dV$ values in the $B-V_{dc}$ plane. For the sample of Fig. 2 (Ja5), we have measured $dI/dV$ traces as a function of $V_{dc}$ at $B$ intervals of 0.2 T and at $T$ = 0.01 K. The colour scale legend on the right hand side shows the various colours used to represent the values of $dI/dV$. The horizontal dashed line denotes $B_c$ (= 0.4 T) of this sample.}
\end{figure}

\end{document}